# Photocurrent Imaging of p-n Junctions and Local Defects in Ambipolar Carbon Nanotube Transistors

Y. H. Ahn,[1] Wei Tsen,[2] Bio Kim,[3] Yung Woo Park,[3] and Jiwoong Park[2,4,*]

1. Division of Energy Systems Research, Ajou University, Suwon 443-749, Korea
2. Department of Chemistry and Chemical Biology, Cornell University, Ithaca NY 14853
3. Department of Physics and Astronomy, Seoul National University, Seoul 151-742, Korea
4. The Rowland Institute at Harvard, Cambridge MA 02142
* E-mail: jp275@cornell.edu

We use scanning photocurrent microscopy (SPCM) to investigate the properties of internal p-n junctions as well as local defects in ambipolar carbon nanotube (CNT) transistors. Our SPCM images show strong signals near metal contacts whose polarity and positions change depending on the gate bias. SPCM images analyzed in conjunction with the overall conductance also indicate the existence and gate-dependent evolution of internal p-n junctions near contacts in the n-type operation regime. To determine the p-n junction position and the depletion width with a nanometer scale resolution, a Gaussian fit was used. We also measure the electric potential profile of CNT devices at different gate biases, which shows that both local defects and induced electric fields can be imaged using the SPCM technique. Our experiment clearly demonstrates that SPCM is a valuable tool for imaging and optimizing electrical and optoelectronic properties of CNT based devices.

In semiconducting nanostructures, including carbon nanotubes (CNTs)[1,2] and nanowires (NWs),[3] the electronic properties at various interfaces, especially metal contacts, play a crucial role, often dominating the overall performance of nanostructure-based devices.[4-7] In addition, the presence and properties of defects can cause various abnormalities in the conductance properties. To fully understand the overall conductance behavior of a nanostructure with high spatial resolution, various scanning probe microscopy (SPM) techniques have been utilized to locally perturb or electrically contact the target structures.[7,8]

More recently, scanning photocurrent microscopy (SPCM) has been successfully applied to study electrical and photoelectric properties of a number of linear nanostructures, including carbon nanotubes[9,10] and semiconducting nanowires.[11,12] The information generated by SPCM includes the band structure near nanostructure/metal interfaces,[12] carrier relaxation dynamics,[13] and detection of local defects.[10]

In this paper, we report SPCM measurements of CNT transistors at different gate biases ($V_G$'s) to investigate the electronic band structure. Our measurements reveal that the peak photocurrent (PC) spots near the metal contacts move with gate bias in the n-type regime. We measure with nanoscale resolution the intensity and polarity of these PC spots in conjunction with DC conductance to investigate the dynamics of the internal *p-n* junctions. We also study the effect of various local potential variations such as naturally existing defects and induced electrical field due to a partial suspension of the CNTs.

Our CNT transistors were fabricated using standard photo- and electron beam lithography techniques with CVD-grown semiconducting CNTs. CNTs were first synthesized directly on a 220 nm thick thermal oxide layer on a conducting Si substrate that is also used as a back gate. Metal evaporation and liftoff were used to define drain and source electrodes. We used either Cr/Au (5/45 nm) or Ti (50 nm) for the electrodes.

Our SPCM setup uses a diffraction-limited laser beam scanned over the CNT devices while the photocurrent is recorded as a function of the position of the laser spot.[12] We also monitor the reflected light intensity simultaneously to determine the absolute position of the laser spot. The schematic of the setup is shown in Fig 1a. To improve the performance we used a lock-in technique with an intensity-modulated laser ($\lambda$ = 781 nm with 20 kHz modulation, unless otherwise noted). The signal-to-noise ratio is dramatically improved this way and we can reliably detect a low-level current signal (< 10 pA) measured with a low light intensity.

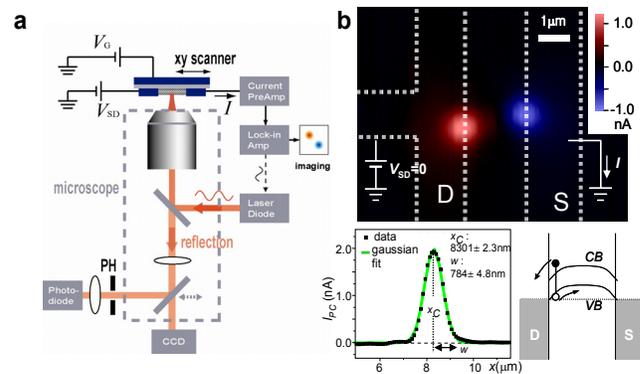

**Figure 1.** (a) Schematic of the SPCM setup (b) SPCM image of a semiconducting CNT device **D1**. Left inset: Gaussian fit to one of the PC spots for determining the peak position and width. Right inset: Mechanism of the contact PC generation

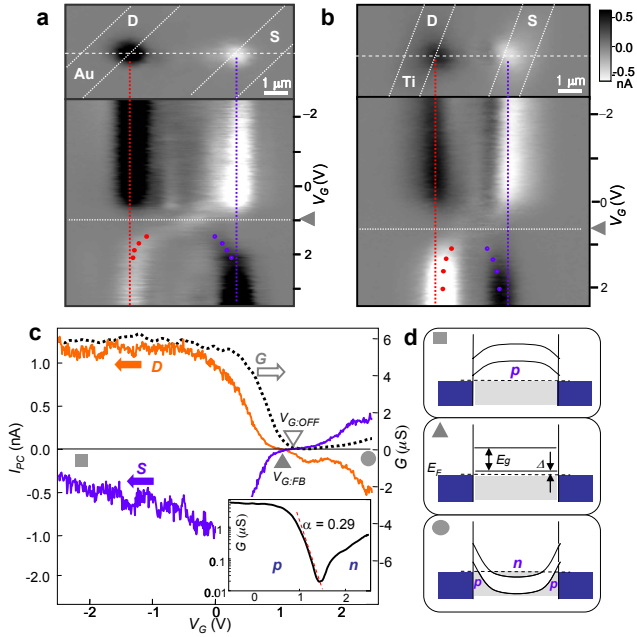

**Figure 2.** Gate bias dependence of PC is shown for semiconducting CNT device **D2** (a) and **D3** (b). (c) PC measured at source (S), drain (D) electrodes and DC conductance (G) are shown. Inset: Semi-logarithmic plot of G vs $V_G$. (d) Band structures for p-type (■), flat band (▲), and n-type (●) cases.

In Fig 1b, we show a representative SPCM image taken at $V_G = V_{SD} = 0$ with a device **D1** fabricated with a semiconducting CNT. It clearly shows localized PC spots near both electrode contacts. These localized PC spots are due to electronic band bending (or equivalently, a local electric field) near the metal/CNT interfaces.[11, 12] Here, the polarity of the PC indicates that the electronic band bends upward toward the middle of the CNT (see the second inset of Fig 1b). Similar photocurrent features were observed from most semiconducting CNT devices measured to date (more than 20 devices), while their magnitude (typically of the order of 1 nA for the light intensity of 100 kW/cm$^2$) varies from device to device.

Since the zero-bias PC signal originates from local electric fields, the position, intensity, and shape of the PC signals provide detailed information about the local electric potential in CNT devices. Furthermore, the peak position and extent of local fields can be determined from the measured PC signal with nanometer-scale resolution (much smaller than the focused laser spot) using a 1D or 2D Gaussian fit method. An example of a 1D Gaussian fit to our PC data is shown in the inset to Fig 1b.

Below we first discuss the properties of the contact PC spots measured as a function of $V_G$ to investigate CNT electron band structure for both p- and n-type operation. To this end, we limit our discussion to CNTs with a relatively small bandgap, the regime where we can clearly observe both p- and n-type operation.

The upper images in Fig 2a and 2b each show a spatial SPCM image of devices **D2** and **D3** measured at a negative $V_G$. The lower images show the gate-dependent behavior of the contact PC spots, measured along the dashed line in the upper image, $V_G$ being swept from -2.5V to a positive value (3.5 V in 2a and 2.5 V in 2b). **D2** and **D3** are contacted by Cr/Au and Ti, respectively. The CNTs in both devices are semiconducting with a relatively small bandgap (estimated to be less than 0.3 eV from the conductance measurements). In both devices, the PC spots near the contacts change polarity at a certain gate bias, $V_{G;FB}$. We also note that in both devices the polarity changes continuously from positive to negative (negative to positive) near the drain (source) electrode as $V_G$ is swept in the positive direction.

The behaviors discussed above are due to the direction change of the electronic band bending at the contact. As $V_G$ increases, the electron band of the middle portion of the CNT moves downward, whereas it is pinned at the contact. As a result, the band-bending direction changes at higher $V_G$. Here the critical gate bias at which the PC polarity flips (zero PC) signifies the configuration where the electron band is flat near either contact (flat band; $V_{G;FB}$). In Fig 2a and 2b, $V_{G;FB}$ is approximately 1.0V and 0.7V respectively, and it is denoted by a solid triangle with a dotted line.

Comparison with the simultaneously measured DC conductance illuminates quantitative details regarding the electron band alignment at the contact. In Fig 2c, we show the PC near the drain (*D*) and source (*S*) electrodes as well as the linear conductance (*G*) as a function of $V_G$ measured from **D2**. Titanium contacted **D3** shows a similar behavior (not shown). The polarity of both PC signals changes around $V_{G;FB} \sim 1.0$ V (denoted by a solid triangle) while the conductance reduces to nearly zero at $V_{G;OFF} \sim 1.1$ V (open triangle). The conductance then increases when $V_G >$ 1.5V as the CNT device becomes *n*-type (see inset, Fig 2c). The conductance shut-off at $V_{G;OFF}$ occurs when the valence band is aligned with the Fermi level of the electrodes. Since $V_{G;FB} < V_{G;OFF}$, the valence band of the CNT is located above the Fermi level of the electrodes at the contact. In **D3**, we observe $V_{G;FB} \sim V_{G;OFF}$ instead, which indicate that the CNT valence band is located close to the electrode Fermi level. In both cases, the energy difference, $\Delta = E_{VB} - E_F$, between the valence band and the Fermi level can be obtained using $\Delta = \alpha|e(V_{G;FB}-V_{G;OFF})|$ where $\alpha$ is the gate efficiency and $e$ is the electron charge. Using the procedure described earlier,[12, 14] we find $\alpha = 0.29$ for **D2** (see inset, Fig 2c), and hence $\Delta \sim 30$ meV. We measure $|\Delta| < 10$ meV for **D3**.

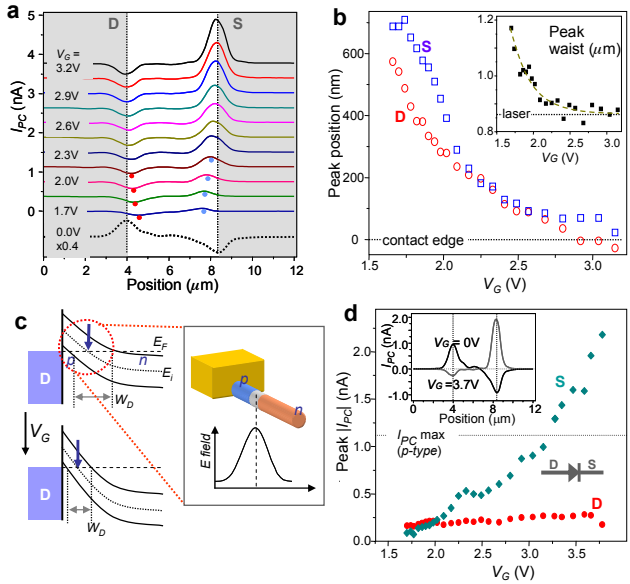

**Figure 3.** Internal *p-n* junction in a *n*-type CNT. (a) PC line scans from **D2** in the *n*-type regime. (b) PC peak position vs $V_G$ obtained using the Gaussian fit method. Inset: Peak waist (drain side) vs $V_G$. (c) A *p-n* junction and depletion region ($W_D$) at different $V_G$'s. (d) PC enhancement and asymmetry in the *n*-type CNT device.

The data shown in Fig 2 together with the discussion given above provide strong evidence that the contact region of the CNT both in **D2** and **D3** is *p*-type. In addition, we expect a large density of holes at the contact in both devices based on the band alignment discussed above. In Fig 2d, the electron band alignment near the contacts in the *p*-type (■), flat band (▲), and *n*-type (●) cases are depicted.

Significantly, in the *n*-type regime there should exist internal *p-n* junctions near the electrodes, as only the main body of the CNT turns *n*-type. The presence of internal *p-n* junctions in CNT devices will have a strong impact both on the conductance and PC behaviors. In particular, a strong local electric field is expected at the *p-n* junction, with its maximum located at the center of the depletion region. Therefore, the maximum PC will coincide with the position of the *p-n* junction. Indeed we observe several distinctive properties of PC signals in the *n*-type regime that are consistent with this model, as described below.

In Fig 3a, we show the line scans of PC measured from **D2** in the *n*-type regime at various $V_G$'s. The most striking feature in this plot is the movement of contact PC spots as $V_G$ changes. The PC peaks move away from the contact region (shaded area) and approach the middle of the CNT as $V_G$ decreases. On the contrary, the PC spots in the *p*-type regime do not change throughout the whole gate bias range.

In Fig 3b, we measure the PC peak positions and widths as a function of $V_G$ using the Gaussian fit method described in Fig 1b, which shows the PC peak movement with the same trend as described above. Surprisingly, the PC peak moves more than 0.6 μm away from the contact, beyond which the PC signal is too weak for a precise measurement. In addition, the Gaussian fit width (inset, Fig 3b) measured from the PC spot indicates an increasing peak width as $V_G$ decreases. Similar gate dependent behaviors were observed in most semiconducting devices studied to date, including **D3** and **D4** (shown in Fig 4).

We can understand these behaviors in terms of the internal *p-n* junctions in *n*-type operation discussed earlier, as shown in Fig 3c. As $V_G$ increases, the main body of the CNT becomes *n*-type when $E_i$ (intrinsic level of the CNT) becomes lower than $E_F$. Since the contact region is still *p*-type, a large portion of the CNT will form a depletion region. As $V_G$ increases further, the position of the depletion region ($W_D$) and the *p-n* junction will move closer to the contact while the width of $W_D$ becomes narrower. The electric field and the potential around the *p-n* junction are analogous to the case of a linearly-varying dopant in conventional semiconductor *p-n* junctions. In this model, the electric field is at a maximum in the middle of the $W_D$ and is strong over the extent of $W_D$.[15]

While the presence of internal *p-n* junctions is also supported by the conductance measurement (this can be seen from the inset to Fig 2c, where the conductance increases exponentially with increasing $V_G$ in the deep *n*-type regime), our data presented in Fig 3 provide a direct measure of the evolution of the *p-n* junctions in much greater spatial detail. Interestingly, we note that the *p-n* junction can stay away from the contact even deep into the *n*-type region, as can be seen for **D3** in Fig 2b (red and blue dots). There, the PC spots stay approximately 200 nm away from the contact throughout most of the *n*-type regime. Previously, low temperature conductance measurements on similar devices reported the formation of a *p*-type quantum dot (size ~ 100 nm) near the contact while the main body of the CNT is *n*-type.[16] Our experiment provides the first photoelectric measurements on such *p-n* junctions with nanoscale spatial resolution.

One of the useful properties of a *p-n* junction is its high efficiency for electron-hole separation and thus sensitive photo-response. We often observe that the *n*-type PC signal is indeed strongly enhanced compared to the *p*-type case as demonstrated in Fig 3d. While the magnitude of *p*-type PC always stays below 1.1 nA, the *n*-type PC near the source contact keeps increasing above 2.3 nA with no saturation. We also observe strong asymmetry between PC signals from source and drain contact. This PC enhancement and asymmetry occurs because one *p-n* junction is more dominant than the other, which makes the entire CNT device equivalent to a single sensitive *p-n* junction photodiode.

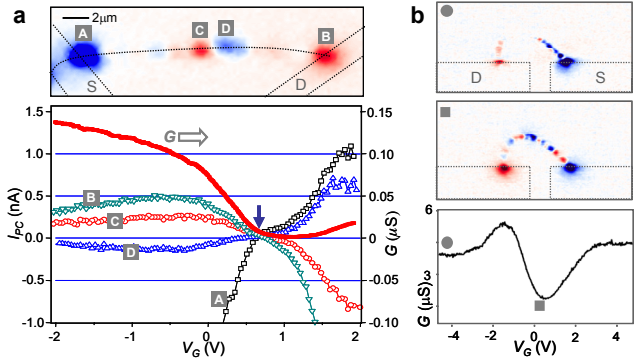
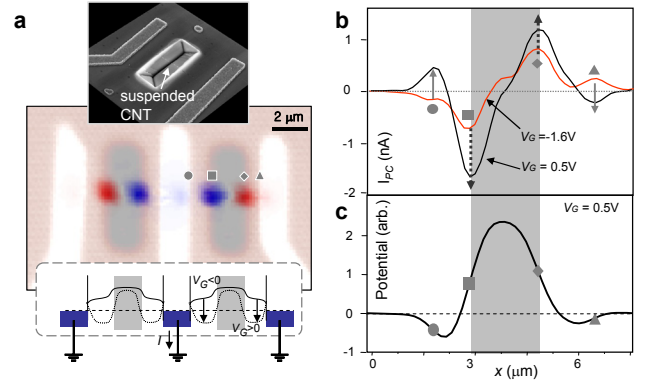

**Figure 4.** PC near CNT defects. (a) SPCM from a 18 μm long CNT device **D4**. Below: PC at points A, B, C, D as well as DC conductance (G) vs $V_G$. (b) SPCM images from **D5** for two different $V_G$'s (measured with λ = 532 nm). The DC conductance vs $V_G$ is shown at the bottom.

**Figure 5.** Electric potential profile in a partially suspended CNT device. (a) An SEM and SPCM image from a suspended CNT device **D6**. Gray areas denote suspended parts. Inset: Expected band movement for different $V_G$'s. (b) PC measured at $V_G$ = -1.6V and 0.5V. (c) Electron energy profile obtained by integrating PC shown in (b).

Since SPCM produces local information regarding the electric field in CNT devices, it can be used to detect and map out potential variations in the device. Below we apply the SPCM technique to investigate two important cases: local defects and suspension-induced local fields.

In Fig 4, we first investigate PC signals associated with defects in CNT transistors. While short CNT devices (**D1-D3**) show strong PC signals only near the contacts, we often observe additional PC spots in the middle of longer CNTs. One representative image for a semiconducting CNT device **D4** (18 μm long) is shown in Fig 4a. One pair of positive and negative PC spots exists in the middle of **D4** in addition to the contact PC spots. A $V_G$-dependent measurement similar to Fig 2a and 2b indicates that these additional spots are localized and do not move with the gate bias, which strongly indicates that a localized defect intrinsic to the CNT device is responsible for this feature.

While the exact nature of this defect (whether it is electrical, mechanical or chemical) is difficult to determine in our experiment, useful information can be obtained by investigating the $V_G$ dependence of this defect induced PC, as shown in the lower image in Fig 4a. The PC measured near the defect (C, D) as well as near the contacts (A, B) is shown together with the linear conductance. Surprisingly, all four PC curves become zero near the conductance shut off (denoted by an arrow) and then change polarity afterwards. Based on the same analysis as in Fig 2, we conclude that the energy level of this defect is pinned near to the Fermi level independent of $V_G$, a behavior that has not been reported previously.

Gate-dependent SPCM images also illuminate the relation between defects/potential variations and DC conductance. In **D5**, shown in Fig. 4b, conductance varies strongly with $V_G$ with no clear bandgap. Zero-bias SPCM images reveal that at $V_G$ = 0 V, where the conductance is strongly suppressed, there are many PC spots in the middle portion of the device. On the contrary, the PC signal continuously evolves from positive to zero, and then to negative at a large negative $V_G$, where the conductance is higher. The same behavior has been observed in other devices with similar conductance features. This suggests that there is a strong correlation between the density of PC spots in the device and the overall conductance. Our recent study with SPM indeed shows that many of the SPCM features are induced by current-limiting local defects. This result will be reported elsewhere.

Finally in Fig. 5, we further demonstrate the use of SPCM to map out the electric potential profile in a partially suspended CNT device, **D6**. Even though this device geometry has attracted wide attention both for nanoelectromechanical systems[17] and high efficiency light emitters,[18] understanding the strength of local electric fields in this device geometry has been difficult due to the limitations of SPM techniques. In Fig 5a, we show both an electron micrograph and a combined SPCM/reflected light image measured on **D6** ($V_G$ = 0V). In this device, a single CNT is contacted by three electrodes, and the middle electrode was used to capture the PC. Strong PC spots are clearly observed at the edges of suspended parts of the CNT, which indicate that the electric field is strong there. This is consistent with the mechanism given for the highly efficient electroluminescenece measured in partially suspended CNT devices.[18] In addition, the polarity of these PC signals indicates that the electron band is higher in the suspended part compared to the supported part (see the inset). The absence of a PC signal near the contacts indicates that the band is flat there.

When we change $V_G$, it is expected that the electron band in the supported parts moves more closely with $V_G$ due to the larger dielectric constant of the substrate ($\varepsilon_{SiOx}$ = 3.8) compared to air. Indeed this is the case as shown in the line scans taken at different $V_G$'s. If we first decrease $V_G$ to -1.6 V, the band of the supported parts move upward to produce significant contact PC signals. As we increase $V_G$ from negative to positive (0.5 V), the electron band of the supported parts moves downward much faster than in the suspended parts. As a result, the PC at the edges of the suspended parts becomes much stronger due to a larger energy difference between adjacent CNT portions and the correspondingly stronger electric field that exists at these locations. In addition, the contact PC changes its polarity at a positive $V_G$. In Fig 5c, we plot the electron band obtained by integrating the PC curve for $V_G$ = 0.5 V, which closely matches the expected band movement shown in the inset to Fig 5a.

In conclusion, the SPCM is a powerful tool for understanding the electron band structure in CNT-based devices. It can produce detailed information regarding the properties of internal *p-n* junctions, local defects, and induced electric fields. Furthermore, the resulting photoelectric information in diverse device configurations with nanoscale spatial resolution will serve as an important guideline for optimizing and interrogating the underlying mechanisms in CNT based electrical, optoelectronic and photovoltaic devices.

We thank M. Burns for helpful discussions. This work was mainly supported by the Rowland Junior Fellow program. YHA is also supported by the Nanoscopia Center of Excellence at Ajou University. Bio Kim was partially supported by the BK21 Frontier Physics Research Division of Seoul National University.